\documentclass[aoas,MSNbibl,nameyear,dvips]{arximspdf}
\usepackage{dcolumn}
\usepackage{graphicx}

\doi{10.1214/14-AOAS730} 
\volume{8}
\issue{2}
\pubyear{2014}
\firstpage{1182}
\lastpage{1208}
\setattribute{copyright}{owner}{In the Public Domain}

\makeatletter
\newcolumntype{d}[1]{D{.}{.}{#1}}
\def\wt{\widetilde}
\def\wh{\widehat}
\def\wh{\widehat}

\def\bone{{\mathbf{1}}}
\def\bSigma{{\bolds{\Sigma}}}
\def\bbeta{{\bolds{\beta}}}

\def\ba{{\mathbf{a}}}
\def\bd{{\mathbf{d}}}

\def\bF{{\mathbf{F}}}
\def\bO{{\mathbf{O}}}
\def\bQ{{\mathbf{Q}}}
\def\bS{{\mathbf{S}}}
\def\bU{{\mathbf{U}}}

\newcommand{\argmin}{\mathop{\operatorname{arg\,min}}}

\newcommand{\argmax}{\mathop{\operatorname{arg\,max}}}

\makeatother

\begin{document}
\begin{frontmatter}

\title{Combining isotonic regression and EM algorithm to predict genetic
risk under monotonicity constraint}
\runtitle{Isotonic regression and EM for genetic risk prediction}
\begin{aug}
\author[a]{\fnms{Jing}~\snm{Qin}\thanksref{m1}\ead[label=e1]{jingqin@niaid.nih.gov}},
\author[b]{\fnms{Tanya~P.}~\snm{Garcia}\thanksref{t2,m2}\ead[label=e2]{tpgarcia@srph.tamhsc.edu}},
\author[c]{\fnms{Yanyuan}~\snm{Ma}\thanksref{m3}\ead[label=e3]{ma@stat.tamu.edu}},
\author[d]{\fnms{Ming-Xin}~\snm{Tang}\thanksref{m4}\ead[label=e4]{mxt1@columbia.edu}},
\author[e]{\fnms{Karen}~\snm{Marder}\thanksref{t5,m4}\ead[label=e5]{ksm1@cumc.columbia.edu}}
\and
\author[d]{\fnms{Yuanjia}~\snm{Wang}\corref{}\ead[label=e6]{yw2016@columbia.edu}\thanksref{t6,m4}}
\runauthor{J. Qin et al.}

\affiliation{National Institute of Allergy and Infectious
Diseases\thanksmark{m1},
Texas A\&M Health University Science Center\thanksmark{m2},
Texas A\&M University\thanksmark{m3}
and\break  Columbia University\thanksmark{m4}}

\thankstext{t2}{Supported by the Huntington's Disease
Society of America, Human Biology Project Fellowship.}
\thankstext{t5}{Supported by NS036630 Parkinson disease Foundation, UL1
RR024156.}
\thankstext{t6}{Supported by NIH Grant NS073671.}

\address[a]{J. Qin\\
Biostatistics Research Branch\\
National Institute\\
\quad of Allergy and Infectious Diseases\\
6700B Rockledge Drive, MSC 7609\\
Bethesda, Maryland 20892-7609\\
USA\\
\printead{e1}}

\address[b]{T.~P. Garcia\\
Department of Epidemiology and Biostatistics\\
Texas A\&M University Health Science Center\\
TAMU 1266\\
College Station, Texas 77843-1266\\
USA\\
\printead{e2}}

\address[c]{Y. Ma\\
Department of Statistics\\
Texas A\&M University\\
TAMU 3143\\
College Station, Texas 77843-3143\\
USA\\
\printead{e3}}

\address[d]{M.-X. Tang\\
Y. Wang\\
Department of Biostatistics\hspace*{77pt}\\
Columbia University\\
630 West 168th Street\\
New York, New York 10032\\
USA\\
\printead{e4}\\
\phantom{E-mail:\ }\printead*{e6}}

\address[e]{K. Marder\\
Department of Neurology\\
Columbia University\\
630 West 168th Street\\
New York, New York 10032\\
USA\\
\printead{e5}}

\end{aug}

\received{\smonth{9} \syear{2013}}
\revised{\smonth{1} \syear{2014}}

%
\begin{abstract}
In certain genetic studies, clinicians and genetic counselors are
interested in estimating the cumulative risk of a disease for
individuals with and without a rare deleterious mutation. 
Estimating the cumulative risk is difficult, however, when the
estimates are based on family history data. 
Often, the genetic mutation status in many family members is
unknown; instead, only estimated probabilities of a patient having a
certain mutation status are available. Also, ages of disease-onset are
subject to right censoring. Existing methods to estimate the cumulative
risk 
using such family-based data only provide estimation at individual time
points, 
and are not guaranteed to be monotonic or nonnegative. In this paper,
we develop a novel method that combines Expectation--Maximization and
isotonic regression to estimate the
cumulative risk across the entire support. Our estimator is monotonic,
satisfies self-consistent estimating equations and has high power in
detecting differences between the
cumulative risks of different populations. 
Application of our estimator to a Parkinson's disease (PD) study
provides the age-at-onset distribution of PD in PARK2 mutation carriers
and noncarriers, and reveals a significant difference between the
distribution in compound heterozygous carriers compared to noncarriers,
but not between heterozygous carriers and noncarriers.
\end{abstract}

%

\end{frontmatter}\newpage

\section{Introduction} \label{sec:intro}
In genetic epidemiology studies [\citet{Struewingetal1997,Marderetal2003,Goldwurm2011}], family history data is collected to
estimate the cumulative distribution function 
of disease onset in populations with different risk factors (e.g.,
genetic mutation carriers and noncarriers).
Such estimates provide crucial information to assist clinicians,
genetic counselors and patients to make important decisions such as
mastectomy [\citet{Gradyetal2013}]. The family history data, however,
raises serious challenges when 
estimating the cumulative risk. First, a family member's exact risk
factor is unknown; the only available information is the estimated
\emph
{probabilities} that a family member has each 
risk factor.
Second, ages of disease onset are subject to censoring due to patient
drop-out or loss to follow-up. For such family history data, the
cumulative risk of disease is 
thus a mixture of cumulative distributions for the 
risk factors with known mixture probabilities.
While different parametric and nonparametric estimators have been
proposed for estimating these mixture data distribution functions, they
are not guaranteed to be monotonic or nonnegative, two principle
features of distribution functions. Most of these estimators also
examine the mixture distributions only at individual time points,
rather than at a range of time points.
To overcome these challenges, we develop a novel, simultaneous
estimation method which combines isotone regression [\citet
{Barlowetal1972}] with an Expectation--Maximization (EM) algorithm. Our
algorithm is based on the binomial likelihood at all observations
[\citet{Huangetal2007,MaWang2013}], and yields estimated distribution
functions that are nonnegative, monotone, consistent, efficient and
that provide estimates of the cumulative risk over a range of time points.


Family history data is often collected when studying the risk of
disease associated with rare mutations
[\citet{Struewingetal1997,Marderetal2003,Wangetal2008,Goldwurm2011}]. 
For example, estimating the probability that Ashkenazi Jewish women
with specific mutations of BRCA1 or BRCA2 will develop breast cancer
[\citet{Struewingetal1997}], estimating the survival function from
relatives of
Huntington's Disease probands with expanded C-A-G repeats in the
huntingtin gene [\citet{Wangetal2012}] and, in this paper, estimating
age at onset of Parkinson's disease in carriers of PARK2 mutations
(Section~\ref{sec:intro-corepd}).

In all these cases, a sample of (usually diseased) subjects referred to
as probands 
are genotyped. Disease history in 
the probands' first-degree relatives, including age at onset of the
disease, is obtained through validated interviews [\citet
{Marderetal2003}]. Because of practical considerations, including high
costs or unwillingness to undergo genetic testing, the relatives'
genotype information is not collected. Instead, the probability that
the relative has the mutation or not is computed based on the
relative's relationship to the proband and the proband's mutation
status [\citet{Khouryetal1993}, Section~8.4]. 
Thus, the distribution of the relative's age at onset of a disease is
a mixture of genotype-specific distributions with known,
subject-specific mixing proportions.


A first attempt at estimating the mixture distribution functions was
based on assuming parametric or semiparametric forms [\citet
{Wuetal2007}] for the underlying mixture densities.
To avoid model misspecification, however, nonparametric estimators such
as the nonparametric maximum likelihood estimator (NPMLE) were also
proposed. 
While in many situations the NPMLEs are consistent and efficient, they
are neither for the mixture model [\citet{Wangetal2012,MaWang2013}].
As improvements over the NPMLEs, 
\citet{Wangetal2012} and
\citet{MaWang2013} proposed consistent and efficient nonparametric
estimators based on estimating equations. 
The estimators stem from casting the problem into a semiparametric
theory framework and identifying the efficient estimator. The resulting
estimator, however, can have computational difficulties when the data
is censored, as it uses inverse probability weighting (IPW) and
augmented IPW to estimate the mixture distribution functions [\citet
{Wangetal2012}].
The weighting function involves a Kaplan--Meier estimator which can
result in unstable estimation because the weighting function can be
close to zero in the right tail. There is
also no guarantee that the resulting estimator is monotonic or
nonnegative, thus, a post-estimate adjustment was implemented to
ensure monotonicity.

In this paper, we propose a novel nonparametric estimator that is
neither complex nor computationally intensive, and yields a genuine
distribution for the mixture data problem under the monotonicity
constraint of a distribution function. 
Providing nonparametric estimators for survival functions under ordered
constraints has received considerable attention recently [\citet
{Parketal2012,Barmi2013}], but the emphasis has been on nonmixture
data. The method we propose is applicable to mixture data. Our method
is motivated from a real-world study on genetic epidemiology of
Parkinson's disease (see Section~\ref{sec:intro-corepd}) and is based
on maximizing a binomial likelihood simultaneously at all observations
[\citet{Huangetal2007}]. 
Our method involves combining an EM 
algorithm
and isotone regression [\citet{Ayeretal1955}] 
so that monotonicity is ensured.
We demonstrate that our estimator is consistent, satisfies
self-consistent estimating equations and yields large power in
detecting differences between the distribution functions in the mixture
populations. Our estimator is easy to implement and, for nonmixture data,
we show that our method coincides with the NPMLE.

\subsection{CORE-PD study to estimate the risk of PARK2 mutations}
\label{sec:intro-corepd}

Parkinson's disease (PD) is a neurodegenerative disorder of the central
nervous system that results in bradykinesia, tremors and problems with
gait. 
PD mostly affects the elderly 50 and older, but early onset cases do
occur and are hypothesized to be a result of genetic risk factors. 
Mutations in the PARK2 gene [\citet{Kitadaetal1998,Hedrichetal2004}]
are the most common genetic risk factor for early-onset PD [\citet
{Luckingetal2000}] and may be a risk factor for late onset [\citet
{Oliveiraetal2003}]. 
While mutations in the PARK2 gene are rare, genetic or acquired defects
in Parkin function may have far-reaching implications for the
understanding and treatment of both familial and sporadic PD.

To understand the effects of mutations in the PARK2 gene, the
Consortium on Risk for Early Onset PD (CORE-PD) study was begun in 2004
[\citet{Marderetal2010}]. 
Experienced neurologists performed in-depth examinations (i.e.,
neurological, cognitive, psychiatric assessments) of proband
participants, a subset of noncarriers, and some of the first-degree
relatives of probands and noncarriers. For relatives who were not
examined in person, their PARK2 genotypes were not available, but their
age at onset of PD was obtained through systematic family history
interviews [\citet{Marderetal2003}]. Based on this family history data,
the objective then is to determine the age-specific cumulative risk of
PD in PARK2 mutation carriers and noncarriers. The results will help
patients interpret a positive test result both in deciding treatment
options and making important life decisions such as family planning.

The remaining sections of this paper are as follows. 
Section~\ref{sec:bin_main_results} describes our proposed estimator
which involves maximizing a binomial log-likelihood with an EM
algorithm. We demonstrate that the ensuing estimator solves a
self-consistent estimating equation and is consistent for complete and
right censored data. 
We demonstrate in Section~\ref{sec:main_results} that we can
reformulate the estimator using a different EM algorithm, for which we
can apply the pool adjacent violators algorithm (PAVA) from isotone
regression to yield a nonnegative and monotonic estimator. We
demonstrate the advantages of our new estimator over current ones
through extensive simulation studies in Section~\ref{sec:simulation_study}. We apply our estimator to the CORE-PD study in
Section~\ref{sec:real_data} and conclude the paper in Section~\ref{sec:conclusion}. Technical details are in the \hyperref[app]{Appendix}, and additional
numerical results are available in the supplementary material [\citet{supp}].

\section{Binomial likelihood estimation}
\label{sec:bin_main_results}

To simplify the presentation, we focus on a mixture distribution with
two components; the techniques presented can be easily extended to more
than two components.

For $i=1,\ldots,n$, we observe a quantitative measure $S_i$ known to
come from one of $p=2$ populations with corresponding distributions
$F_1,F_2$ and densities $dF_1,dF_2$. For example, in the Parkinson's
disease study, $S_i$ is the age of disease onset, $F_1$ is the
distribution for the PARK2 mutation carrier group, and $F_2$ is for the
noncarrier group. The exact population to which $S_i$ belongs is
unknown (i.e., we do not know whether a family member is a mutation
carrier or noncarrier), but one can estimate the probability $q_{ki}$
that $S_i$ was generated from the $k$th population, $k=1,2$. We suppose
the mixture probability $\bQ_i$ has a discrete distribution, denoted as
$p_{\bQ}({\mathbf{q}}_i)$, with finite support ${\mathbf
{u}}_1,\ldots,{\mathbf{u}}_m$. We also
suppose that $q_{1i}+q_{2i}=1$ and, hence, sometimes write
$q_{1i}\equiv\lambda_i$ and $q_{2i}\equiv1-\lambda_i$. In this case,
instead of referring to the discrete distribution $p_{\bQ
}({\mathbf{q}}_i)$, we
simply refer to the distribution of $\lambda_i$, denoted as $\eta
(\lambda_i)$. 
Furthermore, $S_i$ is subject to right-censoring, so we observe
$X_i=\min(S_i,C_i)$, where $C_i$ is a random censoring time independent
of $S_i$. 
We let $G(\cdot)$ denote the survival function of $C_i$ and $dG(\cdot)$
its corresponding density. Last, we let $\Delta_i=I(S_i\leq C_i)$
denote the censoring indicator.

Our objective is to use the independent, identically distributed (i.i.d.)
data $(\bQ_i={\mathbf{q}}_i,X_i=x_i,\Delta_i=\delta_i)$ to form
a nonparametric
estimator of $\bF(t)=\{F_1(t),F_2(t)\}^T$ that is consistent, monotone
on the support of $S_i$ and efficient. Identifiability of $\bF(t)$ is
ensured since the mixture probabilities are assumed known and $\bQ_i$
are not all the same [\citet{Wangetal2007}]. In fact, if $Q_i$ has at
least $k$ distinguished the support points, then the model is identifiable.
To estimate $\bF(t)$, we first consider the nonparametric log-likelihood
\[
\sum_{i=1}^n \log \bigl(p_{\bQ}({
\mathbf{q}}_i)\bigl\{{\mathbf {q}}_i^T
\bd\bF(x_i)G(x_i)\bigr\} ^{\delta_i}\bigl[\bigl\{1-{
\mathbf{q}}_i^T\bF(x_i)\bigr
\}\,dG(x_i)\bigr]^{1-\delta
_i} \bigr).
\]
Because $p_{\bQ}(q_i)$ is independent of the estimation of $\bF(t)$,
and the censoring times are random, the log-likelihood above simplifies to
%
\begin{equation}
\label{eqn:npmle} \sum_{i=1}^n \log\bigl[
\bigl\{{\mathbf{q}}_i^T \bd\bF(x_i)\bigr
\}^{\delta
_i}\bigl\{1-{\mathbf{q}}_i^T\bF
(x_i)\bigr\}^{1-\delta_i}\bigr].
\end{equation}
%
Different maximizations of (\ref{eqn:npmle}) result in the commonly
used NPMLEs (see Appendix \ref{sec:npmle_solution}). Unfortunately, for
the mixture data problem, they turn out to be inconsistent or
inefficient [\citet{MaWang2012}].

\subsection{Motivation for binomial likelihood formulation}
\label{sec:motivation_binlik}
As an improvement over the NPMLEs, we consider a binomial likelihood
estimator. To motivate this estimator, we first consider a nonmixture
model without censoring. That is, we observe independent observations
$S_1,\ldots,S_n$ generated from a common distribution~$F$. Without loss
of generality, we suppose $S_1\leq S_2\leq\cdots\leq S_n$ (i.e., ties
may occur). We demonstrate that, in this setting, the NPMLE and the
binomial likelihood estimator of $F$ are the same. Thus, because the
NPMLE is most efficient in this setting, the binomial likelihood
estimator is as well.

For nonmixture data without censoring, the nonparametric estimator of
$F$ maximizes
\[
\sum_{i=1}^n \log \,dF(s_i)
\]
with respect to $dF(s_i)$ subject to $\sum_{i=1}^n \,dF(s_i)=1$ and
$dF(s_i)\geq0$. From first principles, the maximizer is the well-known
empirical distribution function, $\wh F_n(t)=n^{-1}\sum_{i=1}^nI(s_i\leq t)$.

On the other hand, the empirical distribution function is also the
maximizer of the following binomial log-likelihood. For distinctive
time points $t_1 < t_2< \cdots< t_h$ and each $S_i$, denote a success
if $S_i >t_j$ and a failure if $S_i\leq t_j$ $i=1,\ldots,n$,
$j=1,\ldots,h$. The probability of a success is $\bar F(t_j):=1-F(t_j)$, and the
probability of a failure is $F(t_j)$.
The times $t_1,\ldots,t_h$ can be arbitrary, but are typically chosen
to span the support of the events $S_i$ so as to estimate the
cumulative distribution function over the full support.

Accounting for all possible successes and failures, the binomial
log-likeli\-hood is
\[
\sum_{j=1}^h\sum
_{i=1}^n \bigl\{I(s_i\leq
t_j)\log F(t_j)+I(s_i>t_j)
\log \bar F(t_j)\bigr\}.
\]
Maximizing the above with respect to each $F(t_j)$ and subject to the
monotonic constraint
$F(t_1)\leq F(t_2)\leq\cdots\leq F(t_h)$ gives
\[
\wh{F}_n(t_j)=n^{-1}\sum
_{i=1}^n I(s_i\leq t_j), \qquad   j=1,\ldots,h.
\]
However, this is exactly the empirical distribution function which, by
definition, satisfies the monotonic constraint.

Therefore, in the nonmixture case, maximizing the nonparametric
log-likelihood with respect to $dF$ is equivalent to maximizing the
binomial log-likelihood with respect to $F$ subject to the monotonic constraint
$F(t_1)\leq F(t_2)\leq\cdots\leq F(t_h)$. Because the two estimators are
equivalent and the NPMLE is known to be most efficient, the resulting
binomial likelihood estimator is fully efficient. Motivated by this
result, 
we anticipate that maximizing the binomial log-likelihood may yield
highly efficient estimators in more general mixture models.

\subsection{Binomial likelihood estimator for censored mixture data}
\label{sec:bin_censored}

We now construct a binomial likelihood estimator for mixture data with
censoring. Again, consider arbitrary time points $t_1 < \cdots< t_h$,
such that for each event time $S_i$, a~success occurs if $S_i > t_j$
and a failure if $S_i \leq t_j$, $i=1,\ldots,n$, $j=1,\ldots,h$.
As in Section~\ref{sec:motivation_binlik}, we allow for ties in the
event times $S_i$, and choose times $t_1,\ldots,t_h$ to span the
support of the event times.

Under censoring, we observe $X_i=\min(S_i,C_i)$, which means a success,
$I(S_i>t_j)$, is unobservable for those subjects who are lost to
follow-up before $t_j$. A natural approach then is to view the
unobserved successes as missing data and to use an EM algorithm to
maximize the constructed binomial log-likelihood.

Let $V_{ij}=I(S_i>t_j)$, the unobserved success. For mixture data, when
$V_{ij}$ is observable (i.e., noncensored data), we have that
$P(V_{ij}=1)=\lambda_i \bar F_1(t_j) + (1-\lambda_i) \bar F_2(t_j)$ and
$P(V_{ij}=0)=\lambda_iF_1(t_j)+(1-\lambda_i)F_2(t_j)$, where
$\bar F_k(t_j)=1-F_k(t_j)$, $k=1,2$. Considering all time points
$t_1,\ldots, t_h$, and all possible successes and failures, the
complete data binomial log-likelihood of $\{I(S_i>t_j)\}$, $i=1,\ldots,n$, $j=1,\ldots,h$, is
\begin{eqnarray*}
&&\sum_{j=1}^h\sum
_{i=1}^n\bigl[ I(s_i\leq
t_j)\log\bigl\{\lambda_i F_1(t_j)+(1-
\lambda_i)F_2(t_j)\bigr\}
\\
&&\quad\qquad{}+I(s_i> t_j)\log\bigl\{\lambda_i \bar
F_1(t_j)+(1-\lambda_i)\bar
F_2(t_j)\bigr\}\bigr].
\end{eqnarray*}
If $V_{ij}=I(S_i>t_j)$ were observable, we could estimate
$\bF(t_j)$, $j=1,\ldots,h$, by maximizing the binomial
log-likelihood with respect to $F_1(t_j)$ and $F_2(t_j)$. However,
because $V_{ij}$ is unobservable, we instead 
use an EM
algorithm for maximization. An EM algorithm at a single $t_j$ was given
in \citet{MaWang2013}, but they did not further pursue it. In fact,
\citet{Efron1967} did impute this.

The EM algorithm we propose is an iterative procedure where at the
$b$th step the imputed $V_{ij}$ is
%
\begin{eqnarray}
\label{eqn:w_formula} w_{ij}^{(b)}&=&E\bigl\{I(S_i>t_j)|x_i
\bigr\}
\nonumber
\\[-8pt]
\\[-8pt]
\nonumber
&=&I(x_i>t_j)+(1-\delta_i)I(x_i
\leq t_j)\frac{\lambda_i\bar
F_1^{(b)}(t_j)+(1-\lambda_i)\bar F_2^{(b)}(t_j)} {
\lambda_i\bar F_1^{(b)}(x_i)+(1-\lambda_i)\bar F_2^{(b)}(x_i)},\nonumber
\end{eqnarray}
based on the observed data $X_i=x_i$.
The E-step is then the imputed binomial log-likelihood
%
\begin{eqnarray}
\label{eqn:estep}&& \sum_{j=1}^h\sum
_{i=1}^n\bigl[\bigl(1-w_{ij}^{(b)}
\bigr)\log\bigl\{\lambda _iF_1(t_j)+(1-
\lambda_i)F_2(t_j)\bigr\}
\nonumber
\\[-8pt]
\\[-8pt]
\nonumber
&&\quad\qquad{}+w_{ij}^{(b)}\log\bigl\{\lambda_i\bar
F_1(t_j)+(1-\lambda_i)\bar
F_2(t_j)\bigr\} \bigr].
\end{eqnarray}
The M-step then maximizes the above with respect to $F_1(t_j)$ and
$F_2(t_j)$; specifically, the M-step involves solving
%
\begin{eqnarray}
\label{eqn:bin_esteqn_censor} -\sum_{i=1}^n
\lambda_i\frac{w_{ij}^{(b)}-\lambda_i\bar
F_1(t_j)-(1-\lambda_i)\bar F_2(t_j)} {
\{\lambda_iF_1(t_j)+(1-\lambda_i)F_2(t_j)\}\{\lambda_i \bar
F_1(t)+(1-\lambda_i)\bar F_2(t)\}}&=&0, 
\nonumber
\\[-8pt]
\\[-8pt]
\nonumber
-\sum_{i=1}^n(1-\lambda_i)
\frac{w_{ij}^{(b)}-\lambda_i\bar
F_1(t_j)-(1-\lambda_i)\bar F_2(t_j)} {
\{\lambda_iF_1(t_j)+(1-\lambda_i)F_2(t_j)\}\{\lambda_i \bar
F_1(t)+(1-\lambda_i)\bar F_2(t)\}}&=&0, 
\end{eqnarray}
for $j=1,\ldots,h$. The solution to (\ref{eqn:bin_esteqn_censor}) leads
to the new estimate $F_1^{(b+1)}(t_j)$ and $F_2^{(b+1)}(t_j)$.
Iterating the E- and M-steps until convergence leads to the binomial
likelihood estimator $\wh\bF(t_j)$, $j=1,\ldots,h$, for censored
mixture data. We now make several observations about this proposed estimator.


The estimating equations in (\ref{eqn:bin_esteqn_censor}) are optimally
weighted [\citet{Godambe1960}] and are, in fact, self-consistent
estimating equations [\citet{Efron1967}]. The self-consistency stems
from the imputation procedure of the EM algorithm, analogously to the
work of \citet{Efron1967}. In the special case of right censoring but
no mixture, the above approach has a closed-form solution, which is the
celebrated Kaplan--Meier estimator [\citet{Efron1967}]. In the general
case, it can be shown that the proposed estimator $\wh\bF$ is
consistent. The proof is trivial if $\bF$ takes discrete finite many
values. On the other hand, if $\bF$ is a continuous distribution, one
may use the law of large sample and Kullback--Leibler information
inequality to prove it. Details are given in the Appendix
\ref{sec:consistency_imputed_loglikelihood}.
Asymptotics of $\wh\bF(t_j)$ are much more involved, however, and
require solving a complex integral equation which is impractical.
Hence, inference is usually performed using a Bootstrap approach.


Solving for $\wh\bF(t)$ in practice is also a computationally intensive
task. No closed-form solution to (\ref{eqn:bin_esteqn_censor}) exists,
and ensuring monotonicity and nonnegativity of $\wh\bF(t)$ would
actually require solving (\ref{eqn:bin_esteqn_censor}) subject to the
constraints $F_k(t_1)\leq F_k(t_2)\leq\cdots\leq F_k(t_h)$, $k=1,2$, for
$t_1\leq\cdots\leq t_h$. Such a constraint only further complicates
the already demanding estimation procedure. Still, requiring
monotonicity is essential when the data is censored. Without monotonicity,
the imputed weights $w_{ij}^{(b)}$ may not be in the range $(0,1)$, which
could lead to nonconvergence when solving~(\ref{eqn:bin_esteqn_censor}).
Thus, to ensure monotonicity and avoid the complexities of directly
solving~(\ref{eqn:bin_esteqn_censor}), we now describe another approach
for obtaining the binomial likelihood estimator. 

\section{Genuine nonparametric distribution estimators}
\label{sec:main_results}
To construct a monotone and nonnegative estimator $\wh\bF(t)$ at
times $t_1< \cdots< t_h$, we maximize a binomial log-likelihood using
a combined EM algorithm and 
pool adjacent
violators algorithm (PAVA). Before describing the new method, we first
provide a brief overview of PAVA.

\subsection{Pool adjacent violator algorithm}
Isotone regression [\citet{Barlowetal1972}] is the notion of fitting a
monotone function to a set of observed points $y_1,\ldots,y_n$ in a
plane. Formally, the problem involves finding a vector $\ba
=(a_1,\ldots,a_n)^T$ that minimizes the weighted least squares
\[
\sum_{i=1}^n r_i(y_i-a_i)^2
\]
subject to $a_1\leq\cdots\leq a_n$ for weights $r_i>0$, $i=1,\ldots,n$. The solution to this optimization problem is the so-called max-min
formula [\citet{Barlowetal1972}]:
\[
\wh a_j=\max_{s \leq j}\min
_{t\geq j} \frac{\sum_{h=s}^t
y_hr_h}{\sum_{h=s}^t r_h},\qquad    j=1,\ldots,n.
\]
Rather than solving this max-min formula, the weighted least squares
problem is instead solved using PAVA [\citet{Ayeretal1955,Barlowetal1972}], a simple procedure that yields the solution in $O(n)$
time [\citet{GrotzingerWitzgall1984}]. The history of PAVA, its
computational aspects and a fast implementation in R are discussed in
\citet{Leeuwetal2009}. Variations of PAVA implementation include using
up-and-down blocks [\citet{Kruskal1964}] and recursive partitioning
[\citet{Lussetal2010}]. 

Our idea is to apply PAVA to a variant of our binomial loglikehood and
yield a monotone estimator $\wh\bF(t)$. 
It is important to note that we cannot simply apply PAVA to the
estimator solving (\ref{eqn:bin_esteqn_censor}). The E-step in (\ref
{eqn:estep}) is not in the exponential family, which is a requirement
of PAVA [\citet{Robertsonetal1988}].\vadjust{\goodbreak} Furthermore, applying PAVA to
maximize a binomial log-likelihood has been used in current status data
[\citet{JewellKalbfleisch2004}], but not in the context of mixture data
as we do.

\subsection{PAVA-based binomial likelihood estimator for censored
mixture data}
\label{sec:non_censored_data}
We now modify the construction of the binomial likelihood estimator for
censored mixture data (Section~\ref{sec:bin_censored}) so that PAVA may
be applied. In our earlier construction (Section~\ref{sec:bin_censored}), we viewed the event $I(S_i>t_j)$ as the only
missing data, $i=1,\ldots,n$, $j=1,\ldots,h$. Now, we also consider the
unobserved population membership as missing. Let $L_i$ denote the
unobserved population membership for observation $i$. 

Analogous to the argument in Section~\ref{sec:bin_censored}, we first
consider the ideal situation when $L_i$ and $I(S_i>t_j)$ are
observable. We suppose $L_i=1$ when $S_i$ is generated from $F_1$, and
$L_i=0$ when $S_i$ is generated from $F_2$. In this case,
$P(L_i=1)=\lambda_i$ and $P(L_i=0)=1-\lambda_i$.
For mixture data, the probability $S_i >t_j$ is $\lambda_i \bar
F_1(t_j)$ when $L_i=1$ and is $(1-\lambda_i)\bar F_2(t_j)$ when
$L_i=0$. Likewise, the probability $S_i\leq t_j$ is $\lambda_i
F_1(t_j)$ when $L_i=1$ and is $(1-\lambda_i)F_2(t_j)$ when $L_i=0$.
Therefore, the complete data log-likelihood of $\{L_i, I(S_i\leq t_j)\}, i=1,2\ldots,n, j=1,2,\ldots,h$, is the binomial log-likelihood
\begin{eqnarray*}
\ell_c &=& \sum_{j=1}^h
\sum_{i=1}^n \bigl[L_iI(S_i
\leq t_j)\log\bigl\{\lambda _iF_1(t_j)
\bigr\} +L_iI(S_i>t_j)\log\bigl\{
\lambda_i\bar F_1(t_j)\bigr\}
\\
&&\quad\qquad{}+ (1-L_i)I(S_i\leq t_j)\log\bigl\{(1-
\lambda_i)F_2(t_j)\bigr\}
\\
&&\hspace*{78pt}\qquad{}+(1-L_i)I(S_i>t_j)\log\bigl\{(1-
\lambda_i)\bar F_2(t_j)\bigr\}\bigr].
\end{eqnarray*}
However, neither the population membership $L_i$ nor the event
$I(S_i>t_j)$ are available. Hence, these values must be imputed, and an
EM algorithm will be used for maximization.

At the $b$th step of the EM algorithm, we compute $E\{L_iI(S_i\leq
t_j)|x_i\}= E\{L_i|S_i\leq t_j\}E\{I(S_i\leq t_j)|x_i\}$ and $E\{
L_iI(S_i > t_j)|x_i\}= E\{L_i|\break S_i>t_j\}E\{I(S_i> t_j)|x_i\}$ based on
observed data $X_i=\min(S_i,C_i)$ with $X_i=x_i$. We found earlier that
$E\{I(S_i> t_j)|x_i\}=w_{ij}^{(b)}$ as defined in (\ref
{eqn:w_formula}). Using a similar calculation, we obtain
\begin{eqnarray*}
u_{ij}^{(b)}&\equiv& E(L_i|S_i
\leq t_j)=\frac{\lambda_i
F_1^{(b)}(t_j)}{\lambda_i F_1^{(b)}(t_j)+(1-\lambda
_i)F_2^{(b)}(t_j)},
\\
v_{ij}^{(b)}&\equiv& E(L_i|S_i>
t_j)=\frac{\lambda_i\bar
F_1^{(b)}(t_j)}{\lambda_i\bar F_1^{(b)}(t_j)+(1-\lambda_i)\bar
F_2^{(b)}(t_j)}.
\end{eqnarray*}
Therefore, at the $b$th step, with observed data $\bO^{(b)}=\{X_i\},
i=1,\ldots,n$, the E-step is
\begin{eqnarray*}
E\bigl(\ell_c|\bO^{(b)}\bigr)&=& \sum
_{j=1}^h \sum_{i=1}^n
\bigl[u_{ij}^{(b)}\bigl(1-w_{ij}^{(b)}
\bigr)\log\bigl\{\lambda_i F_1(t_j)\bigr\}
+v_{ij}^{(b)}w_{ij}^{(b)}\log\bigl\{
\lambda_i\bar F_1(t_j)\bigr\}
\\
&&\quad\qquad{}+ \bigl(1-u_{ij}^{(b)}\bigr) \bigl(1-w_{ij}^{(b)}
\bigr)\log\bigl\{(1-\lambda_i)F_2(t_j)\bigr
\}
\\
&&\hspace*{85pt}\qquad{}+\bigl(1-v_{ij}^{(b)}\bigr)w_{ij}^{(b)}
\log\bigl\{(1-\lambda_i)\bar F_2(t_j)\bigr\}
\bigr].
\end{eqnarray*}
The M-step then maximizes the above expression with respect to
$F_1(t_j)$ and $F_2(t_j)$ at each $t_j$. To ensure monotonicity,
however, the M-step actually involves maximizing $E(\ell_c|\bO^{(b)})$
subject to the monotonic constraints $F_k(t_1)\leq F_k(t_2)\leq
\cdots\leq
F_k(t_h)$, $k=1,2$. Though constrained maximization is typically a
challenging procedure, the task is simplified because the
log-likelihood $E(\ell_c|\bO^{(b)})$ belongs to the exponential family,
in which case PAVA is applicable. From the theory of isotonic
regression [\citet{Robertsonetal1988}], we have
\begin{eqnarray*}
&&\argmax_{F_1(t_1)\leq\cdots\leq F_1(t_h)}E\bigl(\ell_c|\bO^{(b)}\bigr)\\
&&\qquad=
\argmin_{F_1(t_1)\leq\cdots\leq F_1(t_h)} \sum_{j=1}^h \sum
_{i=1}^n r_{1ij}^{(b)}
\biggl\{u_{ij}^{(b)}\frac
{1-w_{ij}^{(b)}}{r_{1ij}^{(b)}}-F_1(t_j)
\biggr\}^2,
\\
&&\argmax_{F_2(t_1)\leq\cdots\leq F_2(t_h)}E\bigl(\ell_c|\bO^{(b)}\bigr)\\
&&\qquad=
\argmin_{F_2(t_1)\leq\cdots\leq F_2(t_h)} \sum_{j=1}^h \sum
_{i=1}^n r_{2ij}^{(b)}
\biggl\{ \bigl(1-u_{ij}^{(b)} \bigr)\frac
{1-w_{ij}^{(b)}}{r_{2ij}^{(b)}}-F_2(t_j)
\biggr\}^2,
\end{eqnarray*}
where $r_{1ij}^{(b)}=u_{ij}^{(b)} (1-w_{ij}^{(b)}
)+v_{ij}^{(b)}w_{ij}^{(b)}$
and
$r_{2ij}^{(b)}= (1-u_{ij}^{(b)} ) (1-w_{ij}^{(b)}
)+ (1-v_{ij}^{(b)} )w_{ij}^{(b)}$.

These formulations suggest that $\{F_1(t_j)\}_{j=1}^h$ is the weighted
isotonic regression of $
u_{ij}^{(b)} (1-w_{ij}^{(b)} )/r_{1ij}^{(b)}$ with weights
$r_{1ij}^{(b)}$. Likewise, $\{F_2(t_j)\}_{j=1}^h$ is the weighted
isotonic regression of $
(1-u_{ij}^{(b)} ) (1-w_{ij}^{(b)} )/r_{2ij}^{(b)}$
with weights $r_{2ij}^{(b)}$.
Thus, the max-min results of isotone regression apply and yield
solutions\looseness=1
\begin{eqnarray*}
\wt F_1^{(b+1)}(t_j)&=&\max
_{s\leq j}\min_{t\geq j}\frac{\sum_{h=s}^t\sum_{i=1}^n u_{ih}^{(b)} (1-w_{ih}^{(b)} )}{\sum_{h=s}^t\sum_{i=1}^n  \{u_{ih}^{(b)} (1-w_{ih}^{(b)} )
+v_{ih}^{(b)}w_{ih}^{(b)} \}},
\\
\wt F_2^{(b+1)}(t_j)&=&\max_{s\leq j}
\min_{t\geq j}\frac{\sum_{h=s}^t\sum_{i=1}^n  (1-u_{ih}^{(b)} )
(1-w_{ih}^{(b)} )}{\sum_{h=s}^t\sum_{i=1}^n  \{
(1-u_{ih}^{(b)} ) (1-w_{ih}^{(b)} )
+ (1-v_{ih}^{(b)} )w_{ih}^{(b)} \}}.
\end{eqnarray*}\looseness=0
Rather than solving these max-min formulas, we instead use the PAVA
algorithm implemented in R [\citet{Leeuwetal2009}]. Iterating through
the E- and M-steps with PAVA leads to a genuine estimator of the
mixture distributions. 

For noncensored data (i.e., $\delta_i=1$, $i=1,\ldots,n$),
$w_{ij}^{(b)}$ in (\ref{eqn:w_formula}) simplifies to
$w_{ij}^{(b)}=I(S_i > t_j)$. In this case, the proposed EM algorithm
with PAVA in the M-step remains as stated but with $w_{ij}^{(b)}=I(S_i
> t_j)$ throughout.

Finally, the proposed EM-PAVA algorithm converges to the maximum
likelihood estimate of the binomial likelihood. This follows because
$E(\ell_c|\bO^{(b)})$ belongs to the exponential family and is convex
[\citet{Wu1983}]. Thus, the derived estimator is the unique maximizer
and satisfies the monotonic property of distribution functions.
\subsection{Hypothesis testing}
\label{sec:hypothesis_tests}
For a two-mixture model, one key interest is testing for differences
between the two mixture distributions, that is, testing $H_0\dvtx F_1(t)=F_2(t)$ vs. $H_1\dvtx F_1(t)\neq F_2(t)$ for a finite set of $t$
values or over an entire range. To test this difference, we suggest the
following permutation strategy [\citet{ChurchillDoerge1994}]. For the
data set given, obtain the estimate $\wt\bF^{(0)}(t)$ using the
EM-PAVA algorithm and compute $s^{(0)}=\sup_{t}|\wt F_1^{(0)}(t)-\wt
F_2^{(0)}(t)|$. Then, for $k=1,\ldots,K$, create a permuted sample of
the data by permuting the pairs $(X_i,\delta_i)$ and coupling them with
the mixture proportions ${\mathbf{q}}_1,\ldots,{\mathbf
{q}}_n$. For the $k$th permuted
data set, compute $\wt\bF^{(k)}(t)$ and $s^{(k)}=\sup_{t}|\wt
F_1^{(k)}(t)-\wt F_2^{(k)}(t)|$. Finally, the $p$-value associated with
testing $H_0$ is $\sum_{k=1}^K I(s^{(k)}\geq s^{(0)})/K$. In practice,
we recommend using $K=1000$ permutation data sets. We compare the power
of various tests in Section~\ref{sec:simulation_study}.\eject

\section{Simulation study}
\label{sec:simulation_study}

\subsection{Simulation design}

We performed extensive simulation studies to investigate the
performance of the proposed EM-PAVA algorithm. We report here the
results of three experiments comparing EM-PAVA to existing estimators
in the literature: the type~I NPMLE, type~II NPMLE (see Appendix \ref
{sec:npmle_solution} for the forms of the NPMLEs), and the oracle
efficient augmented inverse probability weighting estimator (Oracle
EFFAIPW) of \citet{Wangetal2012}, Section~3. ``Oracle'' here refers to
the assumption that the underlying density $\bd\bF(t)$ is known exactly
and is not estimated using nonparametric methods. 

The three experiments were designed as follows:
\begin{longlist}[Experiment 1:]
\item[\textit{Experiment} 1:]
$F_1(t)=\{1-\exp(-t)\}/\{1-\exp(-10)\}$ and $F_2(t)=\{1-\exp
(-t/2.8)\}
/\{1-\exp(-10/2.8)\}$ for $0\leq t\leq10$.
\item[\textit{Experiment} 2:]
$F_1(t)=0.8/[1+\exp\{-(t-80)/5\}]$ for $0\leq t\leq100$ and
$F_1(t)=0.678+0.001t$ for $100\leq t\leq300$. $F_2(t)=0.2/[1+\exp\{
-(t-80)/5\}]$ for $0\leq t\leq100$ and $F_2(t)=-0.205+0.004t$ for
$100\leq t\leq300$. Data is generated as specified, however, the
estimation procedure focuses on estimates of $\bF(t)$ for $0\leq t\leq100$.
\item[\textit{Experiment} 3:] 
$F_1(t)=\{1-\exp(-t/4)\}/\{1-\exp(-2.5)\}$
for $0\leq t\leq10$ and $F_2(t)=\{1-\exp(-t/2)\}/\{1-\exp(-2.5)\}$ for
$0\leq t\leq5$.
\end{longlist}
The second experiment is designed to mimic the Parkinson's disease data
in Section~\ref{sec:real_data}. 
In all experiments, we set the random mixture proportion ${\mathbf
{q}}_i=(\lambda
_i,1-\lambda_i)$ to be
one of $m= 4$ vector values:
$(1,0)^T$, $(0.6,0.4)^T$, $(0.2,0.8)^T$ and $(0.16,0.84)^T$. The four
vector values had an equally likely chance of being selected. Our
sample size was 500 and we generated a uniform censoring distribution
to achieve 0\%, 20\% and 40\% censoring rates.

The primary goal of the simulation studies is to compare the
bias, efficiency and power of detecting distribution differences. Bias
and efficiency were evaluated at different $t$ values. First, we
evaluated the pointwise bias, $\wh\bF(t)-\bF_0(t)$, at different $t$
values, where $\bF_0(t)$ denotes the truth. Specifically, we ran 500
Monte Carlo simulations and evaluated the pointwise
bias at $t=1.3$ in Experiment 1 (Table~\ref{table:simu_pointwise}), at
$t=85$ in Experiment 2 (Table~\ref{table:simu_pointwise}), and
at $t=2$ in Experiment 3 (supplementary material, Table~S.1).

\begin{table}[t!]
\tabcolsep=0pt
\caption{Results for Experiment 1 at $t=1.3$ and Experiment 2 at
$t=85$: Bias, empirical standard deviation (emp sd), average estimated standard
deviation (est sd), and 95\% coverage (95\% cov) of estimators at
different censoring rates. Results based on 500 simulations with sample
size $n=500$}
\label{table:simu_pointwise}
\begin{tabular*}{\textwidth}{@{\extracolsep{\fill}}ld{2.4}ccccd{2.4}ccc@{}}
\hline
&\multicolumn{9}{c@{}}{\textbf{Experiment 1}}\\[-6pt]
&\multicolumn{9}{c@{}}{\hrulefill}\\
&\multicolumn{4}{c}{$\bolds{F_1(t)= 0.7275}$}&&\multicolumn{4}{c@{}}{$\bolds{F_2(t)=
0.3822}$}\\[-6pt]
&\multicolumn{4}{c}{\hrulefill}&&\multicolumn{4}{c@{}}{\hrulefill}\\
\textbf{Estimator}&\multicolumn{1}{c}{\textbf{bias}}&\textbf{emp sd}&
\textbf{est sd}& \textbf{95\% cov}&&
\multicolumn{1}{c}{\textbf{bias}}&\textbf{emp sd}&\textbf{est sd}& \multicolumn{1}{c@{}}{\textbf{95\% cov}}\\
\hline
&\multicolumn{9}{c}{Censoring rate${} = {}$0\%}\\
EM-PAVA& 0.0002 & 0.0471 & 0.0440 & 0.9420&& -0.0015 & 0.0438 & 0.0419
& 0.9480\\
Oracle EFFAIPW& 0.0004& 0.0461 & 0.0440 & 0.9520&& -0.0014 & 0.0435&
0.0419 & 0.9480 \\
type~I NPMLE&-0.0159 & 0.1048& 0.0579& 0.9120&& -0.0029 & 0.0804 &
0.0627 & 0.9160\\
type~II NPMLE&-0.0674 & 0.0588 & 0.0329 & 0.5040&& 0.0824& 0.0473 &
0.0288 & 0.2980\\[3pt]
&\multicolumn{9}{c}{Censoring rate${} = {}$20\%}\\
EM-PAVA& 0.0023 & 0.0491 & 0.0456 & 0.9360&&-0.0024 & 0.0445 & 0.0430 &
0.9520\\
Oracle EFFAIPW& 0.0019 & 0.0488 & 0.0454 & 0.9420&& 0.0011 & 0.0447&
0.0432 & 0.9440
\\
type~I NPMLE &-0.0089 & 0.0921 & 0.0588 & 0.9260&&-0.0041 & 0.0835&
0.0644 & 0.9180 \\
type~II NPMLE & -0.0846 & 0.0849 &0.0440 & 0.5720&& 0.0920 & 0.0720 &
0.0393 & 0.3900\\[3pt]
&\multicolumn{9}{c}{Censoring rate${} = {}$40\%}\\
EM-PAVA&0.0022 & 0.0526 & 0.0486 & 0.9420&& -0.0025 & 0.0464 & 0.0456 &
0.9500\\
Oracle EFFAIPW& 0.0057 & 0.0562 & 0.0486 & 0.9220 &&-0.0017 & 0.0508 &
0.0460 & 0.9360 \\
type~I NPMLE &-0.0103 & 0.0981 & 0.0614 & 0.9160&& -0.0061 & 0.0868 &
0.0674 & 0.9120\\
type~II NPMLE &-0.0954 & 0.0952 & 0.0453 & 0.5580&&0.1008 & 0.0854 &
0.0395 & 0.3800\\[6pt]
\end{tabular*}
\begin{tabular*}{\textwidth}{@{\extracolsep{\fill}}ld{2.4}ccccd{2.4}ccc@{}}
\hline
&\multicolumn{9}{c@{}}{\textbf{Experiment 2}}\\[-6pt]
&\multicolumn{9}{c@{}}{\hrulefill}\\
&\multicolumn{4}{c}{$\bolds{F_1(t)= 0.5848}$}&&
\multicolumn{4}{c@{}}{$\bolds{F_2(t)=0.1462}$}\\[-6pt]
&\multicolumn{4}{c}{\hrulefill}&&\multicolumn{4}{c@{}}{\hrulefill}\\
\textbf{Estimator}&\multicolumn{1}{c}{\textbf{bias}}&\textbf{emp sd}&
\textbf{est sd}& \textbf{95\% cov}&&
\multicolumn{1}{c}{\textbf{bias}}&\textbf{emp sd}&\textbf{est sd}& \multicolumn{1}{c@{}}{\textbf{95\% cov}}\\
\hline
&\multicolumn{9}{c}{Censoring rate${} = {}$0\%}\\
EM-PAVA&-0.0009 & 0.0482& 0.0470 & 0.9540&&-0.0037& 0.0398 & 0.0357 &
0.9280\\
Oracle EFFAIPW& -0.0015 & 0.0480 & 0.0472 & 0.9600&& -0.0036 & 0.0403 &
0.0368 & 0.9480\\
type~I NPMLE &-0.0133 & 0.0890 & 0.0597 & 0.9500&& -0.0034 & 0.0659 &
0.0521& 0.8980\\
type~II NPMLE & -0.0872 & 0.0697 & 0.0349& 0.4520&&0.1035 & 0.0532 &
0.0248 & 0.0520\\[3pt]
&\multicolumn{9}{c}{Censoring rate${} = {}$20\%}\\
EM-PAVA& 0.0002 & 0.0548 & 0.0493 & 0.9300&&-0.0013 & 0.0391& 0.0381&
0.9540\\
Oracle EFFAIPW&0.0006 & 0.0548 & 0.0498 & 0.9340&& -0.0015 & 0.0396 &
0.0389 & 0.9640 \\
type~I NPMLE &-0.0078 & 0.0908& 0.0623 & 0.9160&& -0.0030 & 0.0682 &
0.0544 & 0.8860\\
type~II NPMLE &-0.0959 & 0.0792& 0.0437& 0.4800&&0.1086& 0.0695
&0.0353& 0.1160\\[3pt]
&\multicolumn{9}{c}{Censoring rate${} = {}$40\%}\\
EM-PAVA&-0.0016 & 0.0557 & 0.0525 & 0.9320&&-0.0002 & 0.0425 & 0.0401 &
0.9500\\
Oracle EFFAIPW& 0.0009 & 0.0578 & 0.0525 & 0.9380 && -0.0008& 0.0434 &
0.0410 & 0.9560
\\
type~I NPMLE &-0.0111 & 0.0977 & 0.0650 & 0.9100 &&-0.0043 & 0.0711 &
0.0560& 0.8760 \\
type~II NPMLE &-0.1048 & 0.0857 &0.0454 & 0.4740&& 0.1153 & 0.0846 &
0.0361 & 0.1380 \\
\hline
\end{tabular*}
\end{table}
%

\begin{table}[t!]
\tabcolsep=0pt
\caption{Results for Experiments 1 and 2 across a range of time points:
Integrated absolute bias, average pointwise variance, and average 95\%
coverage probabilities of estimators at different censoring rates.
Results based on 500 simulations with sample size $n=500$}
\label{table:simu_full}
\begin{tabular*}{\textwidth}{@{\extracolsep{\fill}}lcccccccc@{}}
\hline
&\multicolumn{8}{c@{}}{\textbf{Censoring rate}}\\[-6pt]
&\multicolumn{8}{c@{}}{\hrulefill}\\
&\multicolumn{2}{c}{\textbf{0\%}}&&\multicolumn{2}{c}{\textbf{20\%}}&&
\multicolumn
{2}{c@{}}{\textbf{40\%}}\\[-6pt]
&\multicolumn{2}{c}{\hrulefill}&&\multicolumn{2}{c}{\hrulefill}&&
\multicolumn
{2}{c@{}}{\hrulefill}\\
\textbf{Estimator}&$\bolds{F_1(t)}$ & $\bolds{F_2(t)}$ & &
$\bolds{F_1(t)}$ & $\bolds{F_2(t)}$& &$\bolds{F_1(t)}$ &
$\bolds{F_2(t)}$\\
\hline
& \multicolumn{8}{c}{Experiment 1}\\
&\multicolumn{8}{c}{Integrated absolute bias$^*$}\\
EM-PAVA& 0.0085 & 0.0065&& 0.0190 & 0.0071&& 0.0327 & 0.0199\\
Oracle EFFAIPW & 0.0040 & 0.0055 && 0.0248 & 0.0232 && 0.0967 & 0.0689\\
type~I NPMLE & 0.1409 & 0.0407&& 0.2276 & 0.1063 && 0.4726 & 0.5084 \\
type~II NPMLE & 0.4290 & 0.2960&& 0.5656 & 0.3332 &&0.7127 & 0.3814 \\[3pt]
&\multicolumn{8}{c}{Average pointwise variance$^*$}\\
EM-PAVA& 0.0009 & 0.0005&& 0.0012 & 0.0006&& 0.0015 & 0.0014 \\
Oracle EFFAIPW & 0.0009 & 0.0005 && 0.0011 & 0.0007 && 0.0016 & 0.0015\\
type~I NPMLE & 0.0010 & 0.0013&& 0.0013 & 0.0017&& 0.0022 & 0.0038\\
type~II NPMLE & 0.0006 & 0.0003 &&0.0013 & 0.0004&& 0.0024 & 0.0009\\[3pt]
&\multicolumn{8}{c}{Average 95\% coverage probabilities$^{\dagger}$}\\
EM-PAVA& 0.9512 & 0.9551&&0.9530 & 0.9518 &&0.9513 & 0.9535\\
Oracle EFFAIPW & 0.9498 & 0.9557&& 0.9535 & 0.9514 && 0.9519 & 0.9445 \\
type~I NPMLE & 0.9471 & 0.9508&& 0.9378 & 0.9344 && 0.9130 & 0.8458\\
type~II NPMLE & 0.3756 & 0.5838 && 0.4234 & 0.5927 && 0.3890 & 0.6760\\[6pt]
& \multicolumn{8}{c}{Experiment 2}\\
&\multicolumn{8}{c}{Integrated absolute bias$^{**}$}\\
EM-PAVA& 0.1372 & 0.0342&& 0.1140 & 0.0307&& 0.1049 & 0.0261 \\
Oracle EFFAIPW & 0.0966 & 0.0266&& 0.1282 & 0.0729&& 0.2704 & 0.1215\\
type~I NPMLE & 0.1097 & 0.0467&& 0.0770 & 0.0574 && 0.0791 & 0.0557\\
type~II NPMLE & 3.7021 & 2.4581 && 3.9157 & 2.4937 && 4.4027 & 2.5877 \\[3pt]
&\multicolumn{8}{c}{Average pointwise variance$^{**}$ 
}\\
EM-PAVA& 0.0011 & 0.0003&& 0.0013 & 0.0003 && 0.0014 & 0.0003 \\
Oracle EFFAIPW & 0.0011 & 0.0003 && 0.0013 & 0.0003 && 0.0015 & 0.0003
\\
type~I NPMLE & 0.0013 & 0.0007&& 0.0016 & 0.0007&& 0.0017 & 0.0008\\
type~II NPMLE & 0.0006 & 0.0001 && 0.0006 & 0.0001&& 0.0007 & 0.0002 \\[3pt]
&\multicolumn{8}{c}{Average 95\% coverage probabilities$^{\dagger
\dagger}$}\\
EM-PAVA& 0.9564 & 0.9495 && 0.9538 & 0.9513&& 0.9552 & 0.9530 \\
Oracle EFFAIPW & 0.9547 & 0.9436&& 0.9518 & 0.9475 && 0.9507 & 0.9467 \\
type~I NPMLE & 0.9556 & 0.9479&& 0.9506 & 0.9492&& 0.9505 & 0.9481 \\
type~II NPMLE & 0.5738 & 0.4737&& 0.5781 & 0.4740&& 0.5504 & 0.4805\\
\hline
\end{tabular*}
\tabnotetext[]{}{$^*$Computed over $(0,10)$ for $F_1(t)$ and
$F_2(t)$. $^{\dagger}$Computed over $(0,4)$ for $F_1(t)$ and over $(0,9)$
for $F_2(t)$. $^{**}$Computed over $(0,100)$ for $F_1(t)$ and $F_2(t)$.
$^{\dagger\dagger}$Computed over $(48,100)$ for $F_1(t)$ and
$F_2(t)$.}
\end{table}

Second, we evaluated the estimators over the entire range of $t$ values
based on results from 500 Monte Carlo simulations; see
Tables~\ref{table:simu_full} and S.2
(supplementary
material). In this case, we evaluated the estimators based on the
integrated absolute bias (IAB), average pointwise variance and average
pointwise 95\% coverage probabilities. The integrated absolute bias
(IAB) is $\int_0^\infty|\bar
F_k(t)-F_{k0}(t)|\,dt$, $k=1,2$, where $\bar F_k(t)$ is the average
estimate over
the 500 data sets and $F_{k0}$ is the truth. In our simulation study,
the integral in the IAB was computed using a Riemann sum evaluated at
50 evenly spaced time points across the entire range [i.e., over $(0,10)$
in Experiments 1 and 3, and over $(0,100)$ in Experiment 2]. The IAB for
$F_2(t)$ in Experiment 3 was computed over $(0,5)$ because it is only
defined on this interval. The average pointwise variance and average
pointwise 95\% coverage probabilities were also computed over 50 time
points evenly spaced across the entire range [i.e., over $(0,10)$ in
Experiments 1 and 3, and over $(0,100)$ in Experiment 2]. Specifically,
for each of the 50 time points, we computed the pointwise variance and
pointwise 95\% coverage probabilities of the 500 data sets. Then, we
reported the average of the 50 pointwise values.

\begin{table}
\tabcolsep=0pt
\caption{Empirical rejection rates for Experiments 1 and 2. Test of
$F_1(t)=F_2(t)$ over the entire time range was performed using a
permutation test with 1000 permutations. Results based on 1000
simulations (for test under $H_0$) and 200 simulations (for test under
$H_1$), with sample size $n=500$ and 40\% censoring (under $H_1$)}
\label{table:power}
\begin{tabular*}{\textwidth}{@{\extracolsep{\fill}}lccccccccc@{}}
\hline
&\multicolumn{9}{c@{}}{\textbf{Nominal levels}}\\[-6pt]
&\multicolumn{9}{c@{}}{\hrulefill}\\
&\multicolumn{4}{c}{\textbf{Under} $\bolds{H_0\dvtx F_1(t)=F_2(t)}$} & &
\multicolumn{4}{c@{}}{\textbf{Under} $\bolds{H_1\dvtx F_1(t)\neq F_2(t)}$}\\[-6pt]
&\multicolumn{4}{c}{\hrulefill} & &
\multicolumn{4}{c@{}}{\hrulefill}\\
\textbf{Estimator}& \textbf{0.01} & \textbf{0.05}&
\textbf{0.10}& \textbf{0.20} & & \textbf{0.01} & \textbf{0.05}& \textbf{0.10}&
\multicolumn{1}{c@{}}{\textbf{0.20}}\\
\hline
&\multicolumn{9}{c}{Experiment 1}\\
EM-PAVA & 0.0120 & 0.0560 & 0.0950 & 0.1920 && 0.9000 & 0.9800 & 0.9900
& 1.0000 \\
Oracle EFFAIPW & 0.0090 & 0.0500 & 0.0900 & 0.1820&& 0.6150 & 0.7950 &
0.8650 & 0.9350 \\
type~I NPMLE & 0.0130 & 0.0550 & 0.1020 & 0.1970 && 0.6200 & 0.7650 &
0.8450 & 0.9000 \\
type~II NPMLE & 0.0060 & 0.0490 & 0.1020 & 0.2020&& 0.4400 & 0.5150 &
0.5550 & 0.5900 \\[6pt]
&\multicolumn{9}{c}{Experiment 2}\\
EM-PAVA & 0.0170 & 0.0551 & 0.1022 & 0.2094 && 0.9950 & 0.9950 & 0.9950
& 1.0000\\
Oracle EFFAIPW & 0.0140 & 0.0600 & 0.1100 & 0.2050
&& 0.9950 & 0.9950 & 0.9950 & 1.0000\\
type~I NPMLE & 0.0080 & 0.0550 & 0.1120 & 0.2100 && 0.9200 & 0.9400 &
0.9500 & 0.9600 \\
type~II NPMLE & 0.0100 & 0.0550 & 0.1120 & 0.2150&& 0.7000 & 0.7300 &
0.7500 & 0.7700 \\
\hline
\end{tabular*}
\end{table}

Third, we evaluated the type~I error rate and power in detecting
differences between $F_1(t)$ and $F_2(t)$ over the entire range of $t$
values. We investigated the type~I error rate under $H_0\dvtx F_1(t)=F_2(t)$ based on 1000 simulations. In this case, we generated
data so that $F_2(t)$ was set to the form of $F_1(t)$ in each
experiment (see the description of Experiments 1, 2 and 3). Everything
else was left unchanged. The type~I error rate was then computed using
the permutation test in Section~\ref{sec:hypothesis_tests} using 1000
permutations. The power was computed based on 200 Monte Carlo
simulations. That is, we tested for differences between $F_1(t)$ and
$F_2(t)$ when $F_1(t), F_2(t)$ were evaluated at 50 time points evenly
spaced across the entire range: over $(0,10)$ in Experiments 1 and 3, and
over $(0,100)$ in Experiment 2. To compute the empirical power under
$H_1\dvtx F_1(t)\neq F_2(t)$, we used the permutation test in Section~\ref{sec:hypothesis_tests} with 1000 permutations. 
Results are in Tables~\ref{table:power} and S.3
(supplementary material).

\subsection{Simulation results}

\begin{figure}

\includegraphics{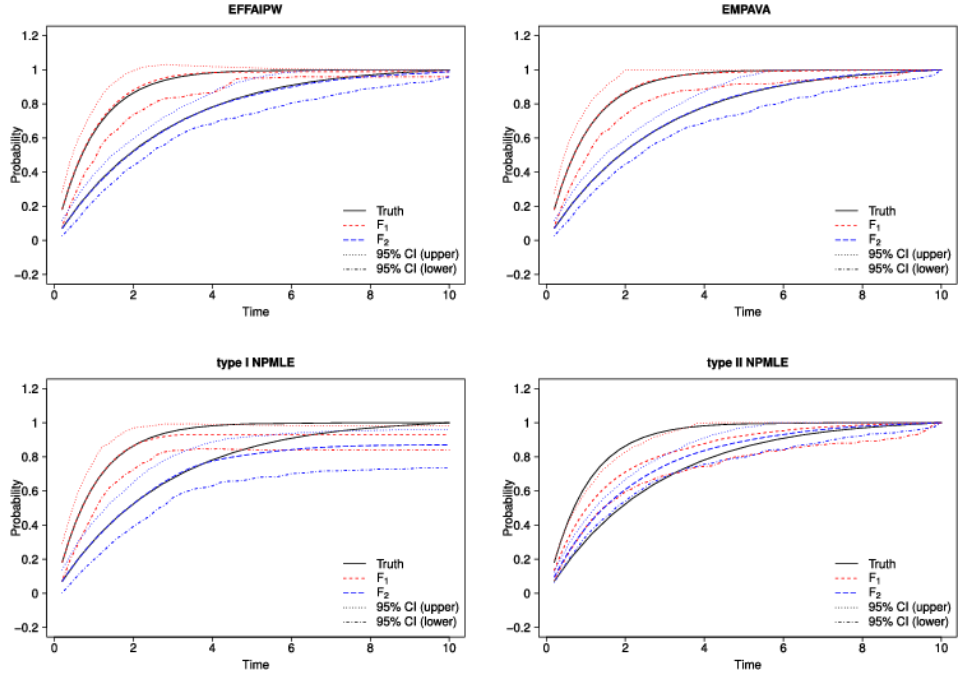}

\caption{Experiment 1. True cumulative distribution function and the
mean of 500 simulations along with 95\% confidence band
(dotted) for the four proposed estimators. Sample size is 500,
censoring rate is 40\%.}
\label{fig:newcox40}
\end{figure}

\begin{figure}

\includegraphics{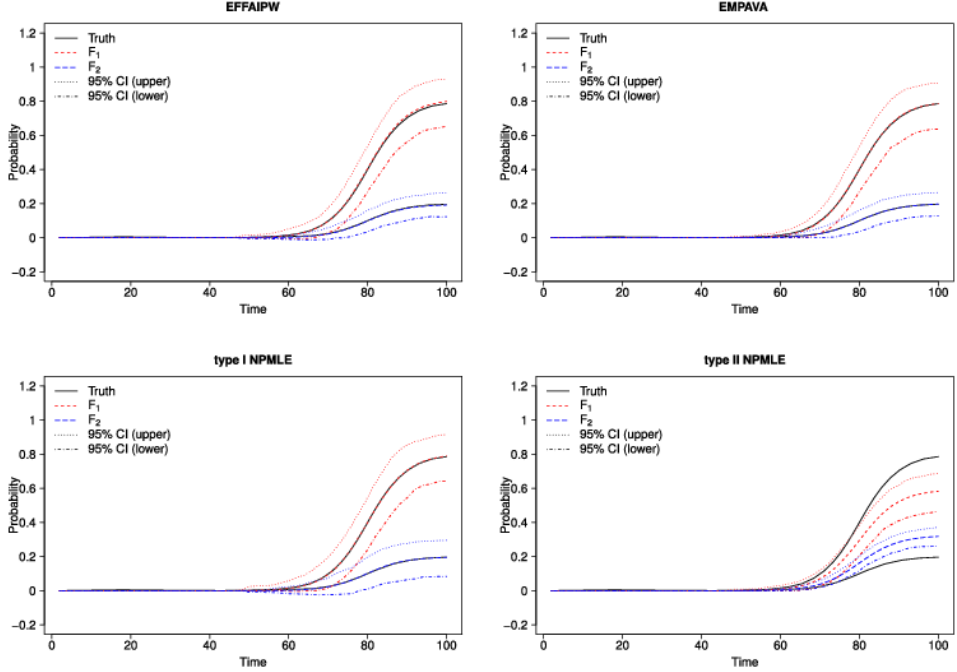}

\caption{Experiment 2. True cumulative distribution function and
the mean of 500 simulations along with 95\% confidence band (dotted)
for the four proposed estimators. Sample size is 500, censoring rate is
40\%.}
\label{fig:cure40}
\end{figure}

Among all four estimators considered, the type~I NPMLE has the largest
estimation variability and the type~II has the largest estimation bias
[see Tables~\ref{table:simu_full} and S.2
(supplementary material)]. In all experiments, as the censoring rate
increases from 0\% to 40\%, the inefficiency for the type~I and the
bias for the type~II worsens. These poor performances alter the 95\%
coverage probabilities, especially for the type~II NPMLE which has
coverage probabilities well under the nominal level (see Table~\ref{table:simu_full}). The inconsistency of the type~II NPMLE is most
apparent in Experiments 1 and 2, where the estimated curve and 95\%
confidence band completely miss the true underlying distributions; see
Figures~\ref{fig:newcox40} and \ref{fig:cure40}. The type~II NPMLE is
also not consistent in Experiment 3, but to a lesser extent; see
Figure~S.1
(supplementary material).

In contrast, across all experiments and censoring rates, the EM-PAVA
estimator performs satisfactorily throughout the entire range of $t$
[see Figures~\ref{fig:newcox40}, \ref{fig:cure40} and S.1
(supplementary material)]. The EM-PAVA estimator is as efficient as the
Oracle EFFAIPW, but with much smaller bias, especially when censoring
is present.
The EM-PAVA also performs well in detecting small differences between
$F_1(t)$ and $F_2(t)$. In Table~\ref{table:power}, the type~I error
rates for all estimators adhere to their nominal levels. When $F_1(t)$
and $F_2(t)$ are largely different (i.e., Experiment 2), then both
EM-PAVA and the Oracle EFFAIPW have similar power in detecting
differences. However, when $F_1(t)$ and $F_2(t)$ are different but to a
lesser degree (i.e., Experiment 1), then EM-PAVA has larger power in
detecting the difference than all other estimators, including the
Oracle EFFAIPW. The larger power of the EM-PAVA estimator is not too
surprising considering that it estimates $\bF(t)$ across a range of
time points, unlike the pointwise estimation of the Oracle EFFAIPW.

A benefit of EM-PAVA over the Oracle EFFAIPW (and the two NPMLEs) is
that EM-PAVA yields a genuine distribution function (i.e., the
estimator is monotone, nonnegative and has values in the $[0,1]$ range).
The curves shown in Figures~\ref{fig:newcox40}, \ref{fig:cure40} and
S.1 (supplementary material) for Oracle EFFAIPW are the
result of doing a post-estimation procedure 
to yield
monotonicity. The ingenuity of the Oracle EFFAIPW estimator, however,
is evident from its 95\% confidence band, which was constructed from
the 2.5\% and 97.5\% pointwise quantiles of the 500 Monte Carlo data
sets. 
Figure~S.1 (supplementary material) shows that the Oracle
EFFAIPW estimator can have 95\% confidence bands outside of the $[0,1]$;
for large $t$ in Figure~S.1, the upper confidence bound is
larger than 1. In contrast, the EM-PAVA estimator is always guaranteed
to be within $[0,1]$ and, thus, its 95\% confidence bands are always
within this range.

\section{Application to the CORE-PD study}
\label{sec:real_data}

\subsection{CORE-PD data and mixture proportions}

We applied our estimator to the CORE-PD study introduced in Section~\ref{sec:intro-corepd}. Data from the CORE-PD study include information
from first-degree relatives (i.e., parents, siblings and children) of
PARK2 probands. The probands had age at onset (AAO) of Parkinson's
disease (PD) less than or equal to 50 and did not carry mutations in
other genes [i.e., neither LRRK2 mutations nor GBA mutations, \citet
{Marderetal2010}].
The key interest is estimating the cumulative risk of PD-onset for the
first-degree relatives belonging to different populations:
\begin{longlist}[1.]
\item[1.]\textit{PARK2 mutation carrier vs. noncarrier}:
We compared the estimated cumulative risk in first-degree relatives
expected to carrying one or more copies of a mutation in the PARK2 gene
(carriers) to relatives expected to carry no mutation (noncarrier).
%
\item[2.]\textit{PARK2 compound heterozygous (or homozygous) mutation
carrier vs. heterozygous mutation carrier vs. noncarrier}: We
considered first-degree relatives who 
have the compound heterozygous genotype (two or more different copies
of the mutation) or homozygous genotype (two or more copies of the same
mutation). We compared distribution of risk in this population to two
different populations: (a) relatives who are expected to have the
heterozygous genotype (mutation on a single allele), and (b) relatives
who are expected to be noncarriers (no mutation). These comparisons
will bring insight into whether heterozygous PARK2 mutations alone
increase the risk of PD or if additional risk alleles play a role. %
\end{longlist}

In the CORE-PD study, the ages at onset for the first-degree relatives
are at least 90\% censored. Information discerning to which population
a relative belongs is available through different mixture proportions.
The mixture proportions are vectors $(p_i, 1-p_i)$, where $p_i$ is the
probability of the $i$th first-degree relative carrying at least one
copy of a mutation. This probability was computed based on the
proband's genotype, a relative's relationship to a proband under the
Mendelian transmission assumption. For example, a child of a
heterozygous carrier proband has a probability of 0.5 to inherit the
mutated allele, and thus a probability of 0.5 to be a carrier. A child
of a homozygous carrier proband has a probability of 1 to be a carrier.
More details are given in \citeauthor{Wangetal2007} (\citeyear{Wangetal2007,Wangetal2008}).
Summary statistics for the populations and the mixture proportions are
listed in Table~\ref{table:summary_data}.

\begin{table}
\tabcolsep=0pt
\caption{Summary statistics for CORE-PD study. Total number of
first-degree relatives ($n$), number of parents, siblings and children,
and percentage of first-degree relatives who have the specified mixture
proportion $(p, 1-p)$, where $p$ is the probability of a relative
carrying at least one copy~of~mutation}
\label{table:summary_data}
\begin{tabular*}{\textwidth}{@{\extracolsep{4in minus 4in}}ld{3.0}d{2.0}d{3.0}
d{3.0}d{3.1}d{2.1}d{1.1}@{}}
\hline
& & & & & \multicolumn{3}{c@{}}{\textbf{Mixture proportion (\%)}}\\[-6pt]
& & & & & \multicolumn{3}{c@{}}{\hrulefill}\\
& \multicolumn{1}{c}{$\bolds{n}$} &
\multicolumn{1}{c}{\textbf{Parents}} &
\multicolumn{1}{c}{\textbf{Siblings}} &
\multicolumn{1}{c}{\textbf{Children}} &
\multicolumn{1}{c}{$\bolds{(1,0)}$} &
\multicolumn{1}{c}{$\bolds{(0,1)}$} &
\multicolumn{1}{c}{$\bolds{(0.5,0.5)}$}\\
\hline
Carrier vs. noncarrier & 355 & 63 & 182 & 110 & 31.5 & 64.8 & 3.7\\
Compound heterozygous  & 17 &1 & 15 & 1 & 100.0 & 0 & 0\\
\quad carrier or homozygous carrier$^*$ &\\
Heterozygous carrier  & 338 & 62 & 167 & 109 & 28.1 &
68.1 & 3.8 \\
\quad vs. noncarrier&&&&&&&\\
\hline
\end{tabular*}
\tabnotetext[]{}{$^*$Genotype
for subjects in this group are known.}
\end{table}

\subsection{Results}

We estimated the cumulative risk based on the EM-PAVA estimator and
compared its results with the type~I NPMLE. The Oracle EFFAIPW\vadjust{\goodbreak}
estimator could not be used because the high censoring led to unstable
estimation: the inverse weights in the estimator were close to zero.
Estimates for the PARK2 compound heterozygous (or homozygous) mutation
carriers were based on a Kaplan--Meier estimator because these subjects
were observed to carry two or more mutations and there is no
uncertainty about the relatives' genotype status (i.e., the data is not
mixture data). We report the cumulative risk estimates along with 95\%
confidence intervals based on 100 Bootstrap replicates.

\begin{figure}

\includegraphics{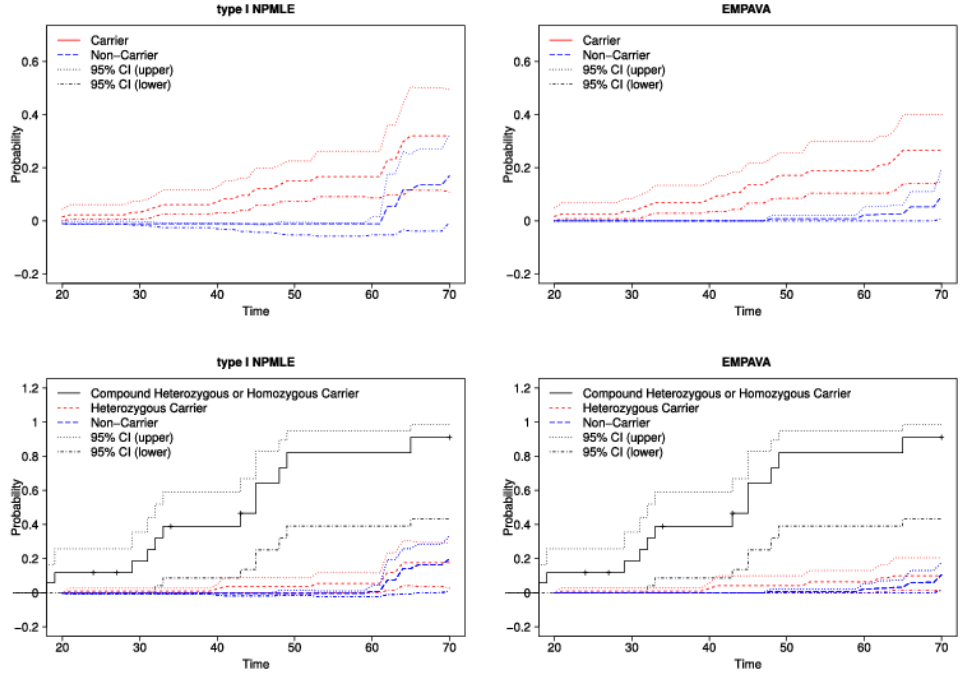}

\caption{CORE-PD study. Estimated cumulative distribution function
for age-at-onset of Parkinson's disease for Parkin mutation carrier
vs. noncarrier (top), and Parkin compound heterozygous or
homozygous carrier vs. Parkin heterozygous carrier and noncarrier (bottom).}
\label{fig:pd_carrier}
\end{figure}

Figure~\ref{fig:pd_carrier} (top right) shows that by age 50, PARK2
mutation carriers have a large increase in cumulative risk of PD onset
compared to noncarriers. Based on EM-PAVA, the cumulative risk (see
Table~\ref{table:real_pointestimates_carrier}) of PD-onset for PARK2
mutation carriers at age 50 is 17.1\% (95\% CI: 8.5\%, 25.6\%), whereas
the cumulative risk for noncarriers at age 50 is 0.8\% (95\% CI: 0\%,
2.1\%). This difference between PARK2 mutation carriers and noncarriers
at age 50 was formally tested using the permutation test in Section~\ref{sec:hypothesis_tests}.
We found that carrying a PARK2 mutation
significantly increases the cumulative risk by age 50
($p$-value${}<0.001$, Table~\ref{table:pvalue}), suggesting that a
mutation in the PARK2 gene substantially increases the chance of early
onset PD. The difference is smaller yet still significant at age 70
($p\mbox{-value }=0.04$, Table~\ref{table:pvalue}).
Even across the age range $(20,70)$, the cumulative risk for PARK2
mutation carriers was significantly different than the cumulative risk
for noncarriers ($p\mbox{-value }=0.01$0, see Table~\ref{table:pvalue}).
These findings are consistent with other clinical and biological
evidence that PARK2 mutations contribute to early-age onset of PD
[\citet
{Hedrichetal2004,Luckingetal2000}].


\begin{table}
\tabcolsep=0pt
\caption{Results for Parkin mutation carriers vs. noncarriers:
Estimated cumulative distribution function and 95\% confidence
intervals (in
parentheses) based on type~I NPMLE and EM-PAVA}
\label{table:real_pointestimates_carrier}
\begin{tabular*}{\textwidth}{@{\extracolsep{\fill}}lccccc@{}}
\hline
& \multicolumn{2}{c}{\textbf{Carrier}} & &
\multicolumn{2}{c@{}}{\textbf{Noncarrier}} \\[-6pt]
& \multicolumn{2}{c}{\hrulefill} & &
\multicolumn{2}{c@{}}{\hrulefill} \\
\textbf{Age} & \textbf{type~I NPMLE} & \textbf{EM-PAVA} &&
\textbf{type~I NPMLE} & \textbf{EM-PAVA}\\
\hline
20 & 0.015 (0.000, 0.043) & 0.017 (0.000, 0.048) &&
$-0.011$ ($-0.009$, 0.000)\phantom{0.} & 0.000 (0.000, 0.000) \\
25 & 0.023 (0.007, 0.061) & 0.026 (0.008, 0.068) &&
$-0.011$ ($-0.013$, $-0.001$) & 0.000 (0.000, 0.000) \\
30 & 0.032 (0.008, 0.073) & 0.036 (0.009, 0.083) &&
$-0.011$ ($-0.016$, $-0.002$) & 0.000 (0.000, 0.000)\\
35 & 0.061 (0.026, 0.116)& 0.068 (0.029, 0.134) &&
$-0.011$ ($-0.026$, $-0.007$) & 0.000 (0.000, 0.000)
\\
40 & 0.072 (0.030, 0.128) & 0.081 (0.034, 0.143) &&
$-0.011$ ($-0.030$, $-0.008$) & 0.000 (0.000, 0.000)\\
45 & 0.121 (0.058, 0.198) & 0.137 (0.067, 0.217) &&
$-0.011$ ($-0.044$, $-0.015$) & 0.000 (0.000, 0.000)\\
50 & 0.150 (0.074, 0.225) & 0.171 (0.085, 0.256) &&
$-0.011$ ($-0.053$, $-0.005$) & 0.008 (0.000, 0.021) \\
55 & 0.166 (0.091, 0.263) & 0.190 (0.104, 0.299) &&
$-0.011$ ($-0.057$, $-0.008$) & 0.008 (0.000, 0.021) \\
60 & 0.166 (0.086, 0.262) & 0.190 (0.105, 0.299) &&
$-0.011$ ($-0.053$, 0.016)\phantom{0.} & 0.023 (0.000, 0.053)\\
65 & 0.321 (0.117, 0.505) & 0.266 (0.138, 0.400) &&
0.117 ($-0.039$, 0.250) & 0.027 (0.000, 0.060) \\
70 & 0.321 (0.109, 0.495) & 0.266 (0.148, 0.400) &&
0.170 ($-0.005$, 0.323) & 0.094 (0.009, 0.193) \\
\hline
\end{tabular*}
\end{table}

\begin{table}
\caption{Results for Parkin compound heterozygous or homozygous carrier
(Compound carrier), Parkin heterozygous carrier and noncarrier:
Estimated cumulative distribution function and 95\% confidence
intervals (in
parentheses)}\label{table:real_pointestimates_splitcar} 
\begin{tabular*}{\textwidth}{@{\extracolsep{\fill}}lcccc@{}}
\hline
\textbf{Age} & \textbf{Kaplan--Meier}$\bolds{^*}$ & &
\textbf{type~I NPMLE} & \textbf{EM-PAVA}
\\
\hline
& {Compound carrier} && \multicolumn{2}{c}{Heterozygous
carrier} \\
20 & 0.118 (0.000, 0.258) && 0.000 (0.000, 0.000)\phantom{0.} & 0.000 (0.000,
0.000)\\
25 & 0.118 (0.000, 0.258) && 0.009 (0.000, 0.027)\phantom{0.} & 0.010 (0.000,
0.030) \\
30 & 0.186 (0.000, 0.355) && 0.009 (0.000, 0.027)\phantom{0.} & 0.010 (0.000,
0.030)\\
35 & 0.389 (0.087, 0.591) && 0.009 (0.000, 0.027)\phantom{0.} & 0.010 (0.000,
0.030) \\
40 & 0.389 (0.087, 0.591) && 0.023 (0.000, 0.049)\phantom{0.} & 0.026 (0.000,
0.056)\\
45 & 0.644 (0.252, 0.830) && 0.037 (0.000, 0.089)\phantom{0.} & 0.041 (0.000,
0.100)\\
50 & 0.822 (0.391, 0.948) && 0.037 ($-0.004$, 0.088) & 0.041 (0.000,
0.100)\\
55 & 0.822 (0.391, 0.948) && 0.056 (0.000, 0.119)\phantom{0.} & 0.063 (0.000,
0.130) \\
60 & 0.822 (0.391, 0.948) && 0.056 ($-0.001$, 0.116) & 0.064 (0.000,
0.131) \\
65 & 0.911 (0.432, 0.986) && 0.177 (0.042, 0.304)\phantom{0.} & 0.100 (0.016,
0.206)\\
70 & 0.911 (0.432, 0.986) && 0.177 (0.027, 0.288)\phantom{0.} & 0.100 (0.016,
0.206) \\[3pt]
&&&\multicolumn{2}{c}{Noncarrier}\\
20 &&& $-0.002$ (0.000, 0.000)\phantom{0.} & 0.000 (0.000, 0.000) \\
25 &&& $-0.002$ ($-0.007$, 0.000) & 0.000 (0.000, 0.000) \\
30 &&& $-0.002$ ($-0.007$, 0.000) & 0.000 (0.000, 0.000) \\
35 &&& $-0.002$ ($-0.007$, 0.000) & 0.000 (0.000, 0.000) \\
40 &&& $-0.002$ ($-0.014$, 0.000) & 0.000 (0.000, 0.000) \\
45 &&& $-0.002$ ($-0.018$, 0.000) & 0.000 (0.000, 0.000) \\
50 &&& $-0.002$ ($-0.015$, 0.014) & 0.008 (0.000, 0.022) \\
55 &&& $-0.002$ ($-0.023$, 0.011) & 0.008 (0.000, 0.022) \\
60 &&& \phantom{0.}0.009 ($-0.022$, 0.044) & 0.023 (0.000, 0.055) \\
65 &&& \phantom{0.}0.142 ($-0.006$, 0.259) & 0.032 (0.000, 0.076) \\
70 &&& 0.199 (0.009, 0.334) & 0.106 (0.015, 0.181) \\
\hline
\end{tabular*}
\tabnotetext[]{}{$^*$Genotype for subjects in this group are known. When there is no
mixture, both methods reduce to Kaplan--Meier.}\vspace*{3pt}
\end{table}

To further distinguish the risk of PD among compound heterozygous or
homozygous carriers (with at least two copies of mutations) from
heterozygous carriers, we separately estimated the distribution
functions in these two groups and compared them to the risk in the
noncarrier group. The numerical results in Table~\ref{table:real_pointestimates_splitcar} and a plot of the cumulative risk
in Figure~\ref{fig:pd_carrier} (bottom panel) indicate a highly
elevated risk in compound heterozygous or homozygous carriers combined.
In contrast, the risk for heterozygous carriers closely resembles the
risk in noncarriers. This result that being a heterozygous carrier has
an essentially similar risk to being a noncarrier was also observed in
another study [\citet{Wangetal2008}]. Further investigation in a larger
study is needed to examine whether risk differs in any subgroup. Using
a permutation test, we also formally tested for differences between the
distribution functions for each group. Results in Table~\ref{table:pvalue} show that there is a significant difference between
compound heterozygous carriers and heterozygous carriers as well as a
significant difference between compound heterozygous and the
noncarriers over the age range $(20,70)$ and at particular ages 50 and
70. Furthermore, there is no significant difference between
heterozygous carriers and noncarriers. These analyses suggest a
recessive mode of inheritance for PARK2 gene mutations for early-onset
PD.


In comparison to the EM-PAVA, the type~I NPMLE had wide and nonmonotone
confidence intervals, which altered the inference conclusions and is
undesirable (see Table~\ref{table:pvalue}). Moreover, the type~I NPMLE
provided a higher cumulative risk in noncarriers by age 70 (17\%),
which appears to be higher than reported in other epidemiological
studies [e.g., \citet{Wangetal2008}]. The poor performance of the type~I NPMLE can be due to instability and inefficiency of the type~I,
especially at the right-tail area. In contrast, EM-PAVA always provided
monotone distribution function estimates, as well as monotone and
narrower confidence bands. The EM-PAVA also gave a lower cumulative
risk in noncarriers by age 70 (9.4\%), which better reflects the
population-based estimates. The increased risk in PARK2 carriers at
earlier ages compared to population-based estimates can also suggest
that there are other genetic and environmental causes of PD in
early-onset cases that are different than late onset.

\begin{table}
\caption{$P$-values associated with testing $H_0\dvtx F_1(t)=F_2(t)$ at
different $t$-values for CORE-PD study. $H_0$ was tested using the
permutation test with 1000 permutations}\label{table:pvalue}
\begin{tabular*}{\textwidth}{@{\extracolsep{4in minus 4in}}lcc@{}}
\hline
& \textbf{type~I NPMLE} & \multicolumn{1}{c@{}}{\textbf{EM-PAVA}}\\
\hline
&\multicolumn{2}{c}{Carrier vs. noncarrier} \\
$t\in[20,70]$ &\phantom{$<$}0.013 & \phantom{$<$}0.010
\\
$t=50$ &$<$0.001 &$<$0.001 \\
$t=70$ &\phantom{$<$}0.073 & \phantom{.}0.04 \\[3pt]
& \multicolumn{2}{c}{Het. carrier vs. noncarrier}\\
$t\in[20,70]$ &\phantom{$<$}0.790 & \phantom{$<$}0.594
\\
$t=50$ & \phantom{$<$}0.341 & \phantom{$<$}0.386 \\
$t=70$& \phantom{$<$}0.813 & \phantom{$<$}0.969 \\[3pt]
&\multicolumn{2}{c}{Compound het./hom. carrier vs. het. carrier} \\
$t\in[20,70]$ & \phantom{$<$}0.013& \phantom{$<$}0.006 \\
$t=50$& $<$0.001& $<$0.001 \\
$t=70$ & \phantom{$<$}0.013& \phantom{$<$}0.017 \\[3pt]
& \multicolumn{2}{c}{Compound het/hom. carrier vs. noncarrier}\\
$t\in[20,70]$& \phantom{$<$}0.011& \phantom{$<$}0.007\\
$t=50$& $<$0.001 & $<$0.001 \\
$t=70$ & \phantom{$<$}0.013& \phantom{$<$}0.017 \\
\hline
\end{tabular*}
\end{table}



\section{Concluding remarks}
\label{sec:conclusion}

In this work we provide nonparametric estimation of age-specific
cumulative risk for mutation carriers and noncarriers. This topic is an
important issue in genetic counseling since clinicians and patients use
risk estimates to guide their decisions on choices of preventive
treatments and planning for the future. For example, individuals with a
family history of Parkinson's disease generally stated that if they
were found to be a carrier and in their mid-thirties, they would most
likely elect to not have children [\citet{Leoetal2005}]. Or, in the
instance they did choose to start a family, PARK2 mutation carriers
were more inclined to undergo prenatal testing [\citet{Leoetal2005}].

It is well known that the NPMLE is the most robust and efficient method
when there is no parametric assumption for the underlying distribution
functions. Unfortunately, in the
mixture model discussed in this paper, the NPMLE (type~II) fails to
produce consistent estimates.
On the other hand, the maximum binomial likelihood method studied in
this paper provides an alternative
consistent estimation method. Moreover, to implement this method, we
have used the combination
of an EM algorithm and PAVA, which leads to genuine distribution
function estimates. For a nonmixture model, the proposed method
coincides with the NPMLE. 
As a result, we expected the proposed method to have high
efficiency, which was apparent through the various simulation studies.
Even though we only considered two-component
mixture models, in principle, the proposed method can be applied to
more than two components
mixture models without essential difficulty.

In some applications, it may be desirable to consider parametric or
semiparametric models (e.g., Cox proportional hazards model,
proportional odds model) in a future work. However, diagnosing model
misspecification has received little attention in the genetics literature.
Our maximum binomial likelihood method can be used as a basis to
construct numerical goodness-of-fit tests. In this case, we can test
whether the distributions conform to a particular parametric or
semiparametric model. That is, the interest is in testing $H_0\dvtx F_1(t)=F_1(t,\bbeta_1), F_2(t)=F_2(t,\bbeta_2)$ for some parametric
models $F_1(t,\bbeta_1)$ and $F_2(t,\bbeta_2)$. To perform this test,
we can use the Kolmogorov--Smirnov goodness of fit
\[
\Delta=\sqrt{n}\max_{-\infty<t< \infty}\bigl\{\bigl|\wt F_1(t)-F_1(t,
\wh \bbeta _1)\bigr|+\bigl|\wt F_2(t)-F_2(t,\wh
\bbeta_2)\bigr|\bigr\},
\]
where $\wh\bbeta_1,\wh\bbeta_2$ are the parametric maximum likelihood
estimates of $\bbeta_1$ and $\bbeta_2$.
Moreover, if one is interested in estimating other quantities of the
underlying distribution functions, for example, the densities, one may
use the kernel method to smooth the estimated
distribution functions.


In our analysis of CORE-PD data, probands were not included due to
concerns of potential ascertainment bias that may be difficult to
adjust [\citet{Begg2002}]. In studies where a clear ascertainment scheme
is implemented, adjustment can be made based on a retrospective likelihood.
Last, the computational procedure of the proposed estimator is simple
and efficient. An R function implementing the proposed method is
available from the authors.



%
%

%
%
%
%

%
%
%
%

\begin{appendix}\label{app}
\section*{Appendix: Sketch of technical arguments}

\subsection{The type~I and type~II NPMLEs}
\label{sec:npmle_solution}
For the type~I NPMLE, let $s_{j}(x_i)={\mathbf{u}}_j^T\bd\bF
(x_i)$ and
$S_{j}(x_i)=1-{\mathbf{u}}_j^T\,\bF(x_i)$, $i=1,\ldots,n$,
$j=1,\ldots,m$. The
type~I NPMLE maximizes
\[
\label{eq:target} \sum_{j=1}^m\sum
_{i=1}^n\log \bigl\{s_j(x_i)^{\delta
_i}S_j(x_i)^{1-\delta
_i}
\bigr\} I({\mathbf{q}}_i={\mathbf{u}}_j)
\]
with respect to $s_j(x_i)$'s and subject to $\sum_{i=1}^n
s_j(x_i)I({\mathbf{q}}_i={\mathbf{u}}_j)\le1$, $s_j(x_i)\ge
0$ for $j=1,\dots,m$.
Because this is equivalent to
$m$ separate maximization problems, each concerning $s_j(\cdot)$ and
$S_j(\cdot)$ only, the maximizers are the classical
Kaplan--Meier estimators:
\[
\wh S_j(t)=\prod_{x_i\le t,{\mathbf{q}}_i={\mathbf{u}}_j} \biggl\{1-
\frac{\delta_i}{\sum_{{\mathbf{q}}_k={\mathbf
{u}}_j}I(x_k\ge x_i)} \biggr\},
\]
with
$s_j(t)=S_j(t^-)-S_j(t)$ for all $t$. With $\wh\bS(t)=\{\wh S_1(t),
\dots, \wh S_m(t)\}^T$ and $\bU=({\mathbf{u}}_1, \dots,
{\mathbf{u}}_m)^T$, the type~I NPMLE is
\[
\wt\bF_{\mathrm{type\ I}}(t) = \bigl(\bU^T\bU \bigr)^{-1}
\bU^T\bigl\{\bone_m-\wh\bS(t)\bigr\}.
\]
Let the variance--covariance matrix of $\wh\bS(t)$ be $\bSigma$, which
is a diagonal matrix
because each of the $m$
components of $\wh\bS(t)$ is estimated using a distinct subset of the
observations. Then, $\wt
\bF_w(t)=(\bU^T\bSigma^{-1}\bU)^{-1}\bU^T\bSigma^{-1}\{\bone
_m-\wh\bS
(t)\}$ is a weighted version of the type~I NPMLE and is more efficient
than the type~I NPMLE.

The type~II NPMLE has no closed-form solution, and an EM algorithm is
typically employed. Specifically, for $k=1,2$, we form at the $b$th
step in the EM algorithm
\[
c_{ik}^{(b)}=\delta_i\frac{
q_{ik}\,dF_k^{(b)}(x_i)
}{\sum_{k=1}^2 q_{ik}
\,dF_k^{(b)}(x_i)}+(1-
\delta_i) \frac{q_{ik}\{1-F_k^{(b)}(x_i)\}
}{\sum_{k=1}^2 q_{ik}\{1-F_k^{(b)}(x_i)\}},
\]
and update the type~II NPMLE estimate as
\begin{eqnarray*}
1-\check F_{{\mathrm{type\ II}},k}^{(b+1)}(t) &=&\prod
_{x_i\le t,\delta_i=1} \biggl\{1-\frac{\sum_{j=1}^n I(x_j=x_i, \delta_j=1)c_{jk}^{(b)}} {
\sum_{j=1}^n c_{jk}^{(b)}I(x_j\ge x_i)
} \biggr\}
\\
&=&\prod_{x_i\le t,\delta_i=1} \biggl\{1-\frac{c_{ik}^{(b)}} {
\sum_{j=1}^n c_{jk}^{(b)}I(x_j\ge x_i)
} \biggr
\}.
\end{eqnarray*}
The procedure is iterated until convergence.

\subsection{Consistency of imputed log-likelihood}
\label{sec:consistency_imputed_loglikelihood}


We first demonstrate consistency for the noncensored data case. When
$\bF$ takes discrete finite many values, the result holds true
trivially. If $\bF$ is a continuous distribution function, then for
noncensored data, the binomial log-likelihood is
\begin{eqnarray*}
\ell&=&\sum_{j=1}^h\sum
_{i=1}^n I(s_i\leq t_j)
\log\bigl[\lambda_i F_1(t_j)+(1-
\lambda_i)F_2(t_j)\bigr]
\\
&&\quad\qquad{}+I(s_i>t_j)\log\bigl[\lambda_i
\bar{F}_1(t_j) +(1-\lambda_i)
\bar{F}_2(t_j)\bigr].
\end{eqnarray*}
This can be written as
\begin{eqnarray*}
n^{-2}\ell&=&\int\int I(s\leq t)\log\bigl[\lambda F_1(t)+(1-
\lambda )F_2(t)\bigr]
\\
&&\qquad{}+I(s>t)\log\bigl[\lambda\bar{F}_1(t) +(1-\lambda)
\bar{F}_2(t)\bigr]\,d\eta_n(s,\lambda)\,d\xi_n(t),
\end{eqnarray*}
where
\[
\eta_n(s,\lambda)=n^{-1}\sum_{i=1}^n
I(s_i\leq s, \lambda_i\leq \lambda ), \qquad   \xi_h(t)=h^{-1}\sum_{i=1}^h
I(t_i\leq t).
\]
By the Law of Large Numbers, it can be shown that
\begin{eqnarray*}
n^{-2}\ell& =& \int\bigl\{\lambda F_{10}(t)+(1-
\lambda)F_{20}(t)\bigr\}\log\bigl\{\lambda F_1(t)+(1-
\lambda)F_2(t)\bigr\}\,d\eta_0(\lambda)\,d
\xi_0(t)
\\
&&{}+ \bigl\{\lambda\bar{F}_{10}(t)+(1-\lambda)\bar{F}_{20}(t)
\bigr\}\log\bigl\{\lambda \bar {F}_1(t)+(1-\lambda)
\bar{F}_2(t)\bigr\}\,d\eta_0(\lambda)\,d\xi
_0(t)\\
&=:&\Delta,
\end{eqnarray*}
where $\eta_0(\lambda)$ is the marginal distribution of $\lambda$ and
\[
\xi_0(t)=\int\bigl\{\lambda F_{10}(t)+(1-
\lambda)F_{20}(t)\bigr\}\,d\eta _0(\lambda).
\]
Here, the subscript $_0$ denotes the truth.
By the Kullback--Leibler information inequality, the above limiting
value achieves the maximum if and
only if $F_1=F_{10}$ and $F_2=F_{20}$. Therefore, the maximum binomial
likelihood
estimation is consistent.

For the censored data case, consistency also holds following a similar argument.
The only difference in the log-likelihood\vadjust{\goodbreak} is that the indicator
function $I(S_i\leq t_j)$
is replaced by $w_{ij}=E\{I(S_i\geq t_j)|S_i\geq x_j\}$. If $\hat
{w}_i(t_j)$ is replaced by
an initial consistency estimation, then the log-censored binomial
likelihood will converge
to~$\Delta$ again.
\end{appendix}
%
%
%
%


\section*{Acknowledgments}
J. Qin and T.~P. Garcia contributed equally to this work.

\begin{supplement}[id=suppA]
\stitle{Additional simulation results}
\slink[doi]{10.1214/14-AOAS730SUPP} 
\sdatatype{.pdf}
\sfilename{aoas730\_supp.pdf}
\sdescription{The supplementary material contains additional simulation
results.}
\end{supplement}



\printaddresses

\end{document}